\documentstyle[prl,aps]{revtex}

\begin{document}

\draft

\title{Quantum corrections to the geodesic equation}

\author{Diego A.\ R.\ Dalvit $^{1}$ \thanks{dalvit@t6-serv.lanl.gov} and
Francisco D.\ Mazzitelli $^{2}$ \thanks{fmazzi@df.uba.ar}}

\address{$^1$ Theoretical Astrophysics, MS B288, Los Alamos National
Laboratory, Los Alamos, NM 87545, USA}

\address{$^2$ 
Departamento de F\'\i sica {\it J. J. Giambiagi}, 
Facultad de Ciencias Exactas y Naturales\\ 
Universidad de Buenos Aires- Ciudad Universitaria, Pabell\' on I\\ 
1428 Buenos Aires, Argentina}

\maketitle

\begin{abstract}
In this talk we will argue that, when gravitons are taken into 
account, the solution to the semiclassical Einstein
equations (SEE) is not physical. 
The reason is simple: any classical device
used to measure the spacetime geometry will also feel the graviton 
fluctuations. As the coupling between the classical device
and the metric is non linear, the device will not measure the `background
geometry' (i.e. the geometry that solves the SEE). As a particular
example we will show that 
a classical particle does not follow a geodesic  of the background
metric. Instead its motion is determined by  a quantum corrected 
geodesic equation that takes into account its coupling to the gravitons.
This analysis will also lead us to find a solution to the so-called
gauge fixing problem: the quantum
corrected geodesic equation is 
explicitly independent of any gauge fixing parameter.
\end{abstract}

\vskip1cm

Talk presented at the meeting {\it Trends in theoretical physics II},
Buenos Aires, Argentina, December 1998

\vskip1cm

\section{Introduction}

In quantum field theory there are many physical situations where one is
interested in the dynamical evolution of fields rather 
than in S-matrix elements. The effective action (EA) is a useful tool to
obtain the equations that govern such dynamics including the backreaction
effects due to quantum fluctuations. However, there are two important problems
that should be solved before one can get meaningful equations.

On the one hand, when the usual effective action is used to derive  
evolution equations, these turn out to be neither real nor causal. The 
cause is that the EA gives evolution equations for ``in-out'' matrix 
elements of the background fields. In order to obtain real and causal 
equations for expectation values, a 
different EA (``in-in'' EA) has 
been introduced, which permits a correct approach to initial value problems
\cite{CTP}. 
On the other hand, both the in-out and the in-in effective actions are 
not physical quantities off-shell.
This is most easily seen in the context of gauge theories, where the EA 
depends on the gauge fixing condition. The scattering matrix is constructed 
going on-shell, and therefore it does not suffer from this problem. The 
equations of motion, on the contrary, are obtained from the off-shell EA,
and are thus gauge fixing dependent.   
The standard approach to tackle this problem is to consider the 
Vilkovisky-DeWitt effective action \cite{V1}, 
which is specifically built to give a reparametrization,
gauge invariant action. However, this action suffers from another type of
arbitrariness, namely the dependence on the supermetric in the space of fields
that is introduced in its definition \cite{Odintsov,KM}. 

Backreaction effects on the spacetime metric are
relevant in different physical situations
like gravitational collapse and black hole evaporation. Any discussion 
of the backreaction problem should include the effect of gravitons which 
contribute to the one loop effective stress tensor with terms of the same 
order as those coming from ordinary matter fields \cite{BD}. 
When graviton loops are included, the metric $g_{\mu\nu}$ that solves the
semiclassical Einstein equations depends on the gauge fixing, and as such it
is not physical. As an example we can mention calculations of compactification
radii in Kaluza-Klein theories \cite{KL}. 
In view of the dependence of the results 
on the gauge fixing, people turned to the Vilkovisky-DeWitt EA as a way
to overcome this setback \cite{KK_VDW}. However it was eventually shown that 
this approach was also incomplete because the results depend on the 
supermetric for the fields manifold \cite{Odintsov}.

In this talk we put forward a solution different from that
advocated in the Vilkovisky-DeWitt EA. Our point is 
that, due to its interaction with the quantum fluctuations of the gravitational
field, a test particle will not follow a $g_{\mu\nu}$-geodesic. Instead its
motion is governed by a quantum corrected geodesic equation, 
which must be gauge 
fixing independent. {\it Therefore, the  solution of the backreaction 
problem consists of two 
steps: to solve the semiclassical Einstein equations and to 
extract the physical quantities from the solution. It is this second 
point that, to our knowledge, has never been considered before.}

In order to illustrate these facts we will consider the calculation
of the leading quantum corrections to the Newtonian potential. As has been 
pointed out in \cite{D1,D2}, when General Relativity is looked upon as an
effective field theory, low energy quantum effects can be studied without the
knowledge of the (unknown) high energy physics. The leading long distance 
quantum corrections to the gravitational interactions are due to massless 
particles and only involve their coupling at energies low compared to the
Planck mass. Using this idea, many authors have calculated the leading 
quantum corrections to the Newtonian potential computing different sets of
Feynman diagrams \cite{D1,D2,VM,HL}. Instead of evaluating diagrams
and S-matrix elements, we are 
here concerned with a covariant calculation based on the effective
action and effective field equations. This covariant approach is more
adequate to study many interesting problems in which
one considers fluctuations around non-flat backgrounds, like 
black hole evaporation, gravitational collapse and
backreaction in cosmological settings, among others. 
We shall first compute the semiclassical Einstein equations for the 
backreaction problem starting from the standard EA and show how they depend on 
the gauge fixing. Using a 
corrected geodesic equation we will deduce a physical quantum corrected 
Newtonian potential, which does not depend on any gauge fixing parameter.
We will also discuss briefly the quantum corrections to the geodesic equation 
in 
a cosmological context.

\section{The Semiclassical Einstein Equations}

The action for gravity coupled to a heavy particle 
(a classical source) has the form \cite{SIGNA}

\begin{equation}
S_G + S_M = \frac{2}{\kappa^2} \int d^4x \sqrt{-{\bar g}} {\bar R} - 
    M \int \sqrt{-{\bar g}_{\mu\nu}dx^{\mu} dx^{\nu} } , 
\label{eq:action}
\end{equation}
where ${\bar R}$ is the curvature scalar,
${\bar g}_{\mu\nu}$ is the metric tensor,
${\bar g}={\rm det} {\bar g}_{\mu\nu}$, and $\kappa^2=32 \pi G$, with
$G$ being Newton's constant. In the background field method quantum 
fluctuations of the gravitational field may be expanded around a 
background metric, ${\bar g}_{\mu\nu}=g_{\mu\nu} + \kappa s_{\mu\nu}$
and a function $\chi^{\mu}$ is chosen to fix the gauge,
which is implemented through a gauge fixing action

\begin{equation}
S_{{\rm gf}} = -\frac{1}{2 \kappa^2} \int d^4x \sqrt{-g} g_{\mu\nu} \chi^{\mu} 
\chi^{\nu} .
\end{equation}

We shall consider the one-parameter family of gauge fixing functions,
the so-called $\lambda$ family,

\begin{equation}
\chi^{\mu}(\lambda) = \frac{1}{\sqrt{1+\lambda}} \left[
g^{\mu\gamma} \nabla^{\sigma} s_{\gamma\sigma} - 
\frac{1}{2} g^{\gamma\sigma} \nabla^{\mu} s_{\gamma\sigma} \right] .
\end{equation} 

For gauge fixing functions linear in the metric fluctuations, ghosts decouple
from the fluctuations $s_{\mu\nu}$ and only couple to the background fields.
The one loop effective action for the background metric is obtained from 
integrating out quantum fluctuations and implies the evaluation of 
functional determinants for gravitons and ghosts in the presence of the
background fields. For the pure gravitational action $S_G$, the one loop 
divergencies in the DeWitt gauge $\lambda=0$ have been calculated long ago 
using dimensional regularization and turn out to be local terms 
quadratic in the curvature tensors \cite{THV}. They read

\begin{equation}
\Delta S_G^{{\rm div}} = \frac{2}{(4-d)96 \pi^2} \int d^4x \sqrt{-g} 
\left[ \frac{21}{10} R_{\mu\nu} R^{\mu\nu} +
       \frac{1}{20} R^2 \right] ,
\label{eq:divergencies}
\end{equation}
where we have omitted the Gauss-Bonnet term, which is a topological invariant
in $d=4$ spacetime dimensions. Apart from the local parts, the one loop EA 
also has non-local components. These have been computed up to
quadratic order in the curvature tensors through a resummation procedure of
the Schwinger DeWitt expansion for the action \cite{Gospel}. In what follows
we shall be working to order ${\cal O}(R^3)$ at the level of the action and
to order ${\cal O}(R^2)$ in the equations of motion. The non-local,
non-analytic terms proportional to $\ln(-\Box)$ are the relevant ones in order
to compute the leading quantum corrections. They can be read off from the 
divergencies in Eq.(\ref{eq:divergencies}) in a manner outlined in 
\cite{D1,Gospel}
\begin{equation}
\Delta S_G^{{\rm nl}} = - \frac{1}{96 \pi^2} \int d^4x \sqrt{-g} 
\left[  \frac{21}{10} R_{\mu\nu} \ln(-\Box) R^{\mu\nu} + 
\frac{1}{20} R \ln(-\Box) R \right] .
\end{equation}

The second term in Eq.(\ref{eq:action}) introduces 
an additional contribution to the EA. Following the method
described in \cite{BV2}, one can compute the new divergence,
from which it is possible to find the  $\ln(-\Box)$
part of the EA arising from the presence of the mass $M$.
After a long calculation we get

\begin{equation}
\Delta S_M^{{\rm nl}} = -\frac{1}{64 \pi^2} \int d^4x \sqrt{-g}
\left[ M_{\mu\nu\rho\sigma} \ln(-\Box)  M^{\rho\sigma\mu\nu} +
2  M_{\mu\nu\rho\sigma} \ln(-\Box) 
   \left( P^{\rho\sigma\mu\nu} -
       \frac{1}{6} R \delta^{\rho(\mu} \delta^{\sigma\nu)} \right) \right] ,
\end{equation}
where 
 \begin{equation}
M_{\mu\nu\rho\sigma}(y) =\frac{M \kappa^2}{8} \int d\tau \delta^4(y-x(\tau))
 \times 
[g_{\mu\nu} {\dot x}_{\rho} {\dot x}_{\sigma} + 
2 {\dot x}_{\mu} {\dot x}_{\nu} {\dot x}_{\rho} {\dot x}_{\sigma}] , 
\label{eq:meff}
\end{equation} 
and
\begin{equation}
P^{\rho\sigma\mu\nu}= 2 R^{\rho(\mu\sigma\nu)} + 
2 \delta^{(\rho[\mu} R^{\sigma)\nu]} - g^{\rho\sigma} R^{\mu\nu} - 
g^{\mu\nu} R^{\rho\sigma} - R \delta^{\rho(\mu} \delta^{\sigma\nu)} 
+ \frac{1}{2} g^{\rho\sigma} g^{\mu\nu} R .
\end{equation}
Here indeces in parenthesis or brackets imply symmetrization with a $1/2$
factor.

As we will calculate long distance corrections to the
Newtonian potential, we can assume that 
the mass $M$ is a classical static 
``point mass'',
although its size should be much larger than its Schwarzschild
radius and the Planck length, in order to justify the weak field
approximation to be done in what follows. 
Its contribution to the nonlocal part of the EA is

\begin{equation}
\Delta S_M^{{\rm nl}} = \frac{7 M \kappa^2}{1536 \pi^2} \int d^4x 
\sqrt{-g} R \ln(-\nabla^2) \delta^3({\vec x}) ,
\end{equation}
where the nonlocal operator $\ln(-\nabla^2)$ acts on the delta function
as $\ln(-\nabla^2) \delta^3({\vec x}) = -\frac{1}{2\pi r^3}$ \cite{DM}.
Adding the classical and quantum contributions of the EA and taking functional
derivations with respect to the metric, it is straightforward
to compute the semiclassical Einstein equations including backreaction of 
gravitons. They can be derived from the in-in EA, or
by taking twice the real and causal part of the in-out equations of motion.
Up to linear order in curvatures they are

\begin{eqnarray}
\frac{1}{8 \pi G} (R_{\mu\nu} - \frac{1}{2} R g_{\mu\nu}) &=&
M \delta_{\mu}^0 \delta_{\nu}^0 \delta^3({\vec x}) - 
\frac{1}{96 \pi^2} \left[ \frac{21}{10} \ln(-\nabla^2) H_{\mu\nu}^{(2)} +
                          \frac{1}{20} \ln(-\nabla^2) H_{\mu\nu}^{(1)} \right]
+ \nonumber \\
&& ~~~~~~~~~~~~~~~~~~~
\frac{7 M \kappa^2}{768 \pi^2} (\nabla_{\mu} \nabla_{\nu} - g_{\mu\nu} 
\nabla^2) \ln(-\nabla^2) \delta^3({\vec x}) ,
\label{see}
\end{eqnarray}
where we have introduced the tensors 
$H_{\mu\nu}^{(1)}=4\nabla_{\mu} \nabla_{\nu} R-4g_{\mu\nu} \nabla^2 R$ 
and 
$H_{\mu\nu}^{(2)}=2\nabla_{\mu} \nabla_{\nu} R-g_{\mu\nu} \nabla^2 R -
2 \nabla^2 R_{\mu\nu}$. Here we have used the fact that the mass $M$ is static
to replace $\Box \rightarrow \nabla^2$. 

In order to solve these equations for the background metric we 
shall make perturbations around flat spacetime,
$g_{\mu\nu}=\eta_{\mu\nu}+h_{\mu\nu}$ with $\eta_{\mu\nu}={\rm diag}(-+++)$. 
We choose the harmonic gauge for the background perturbation metric. 
It is worth  mentioning that this choice is completely independent of 
the gauge fixing problem for the quantum fluctuations. The $00$ component 
for the perturbation $h_{\mu\nu}$ turns out to be

\begin{equation}
h_{00}(\lambda=0) = \frac{2 G M}{r} \left[ 1 + \frac{43 G}{30 \pi r^2} - 
\frac{7 G}{12 \pi r^2} \right] .
\label{eq:DWmetric}
\end{equation}

The first term is due to the
presence of the classical mass $M$ (for simplicity we consider
only the Newtonian limit, that is, we do not include 
classical corrections from general relativity). The second and third
terms are quantum corrections.
The former stems
pure gravitational contributions (vacuum polarization) while the latter
arises from the coupling of the mass $M$ to gravitons.

The above result is valid only for the DeWitt $\lambda=0$ gauge. For any
other gauge (not only for the $\lambda$ family of gauge fixings) 
one has to add a new contribution to the nonlocal part, which should
vanish on-shell. Keeping up to quadratic order in curvatures, the requirement 
that the effective action be gauge fixing independent on-shell fixes the most
general form such a nonlocal term can have

\begin{equation}
\Delta S = \int d^4x \sqrt{-g} 
\left[ a R_{\mu\nu} \ln(-\Box) {\cal E}^{\mu\nu} + 
b R g_{\mu\nu} \ln(-\Box) {\cal E}^{\mu\nu} + 
{\cal O}( {\cal E}_{\mu\nu}^2 ) \right] .   
\label{eq:extra}
\end{equation}

Here ${\cal E}^{\mu\nu}$ is the classical extremal
${\cal E}^{\mu\nu}=-\frac{2}{\kappa^2} (R^{\mu\nu}-\frac{1}{2} R g^{\mu\nu})+
\frac{1}{2} T^{\mu\nu}$, where  
\begin{equation}
T^{\mu\nu}(y) = M \int d\tau \, {\dot x}^{\mu} {\dot x}^{\nu} 
\delta^4(y-x(\tau)) ,
\label{eq:tmunu}
\end{equation}
is the energy-momentum tensor of the classical source
and $a$ and $b$ are constants that depend on which
particular gauge is used. The reason for omitting terms quadratic in the
extremal in Eq. (\ref{eq:extra}) is that, when the equations of 
motion are perturbatively solved, they
vanish identically. This new contribution to the EA modifies the
semiclassical equations (\ref{see}), which will now depend on
$a$ and $b$. Solving the modified equations we obtain
the metric in a general gauge

\begin{equation}
h_{00}= \frac{2 G M}{r} \left[ 1 + \frac{43 G}{30 \pi r^2} - 
\frac{7 G}{12 \pi r^2} +\frac{a - 2b}{r^2} \right] .
\label{eq:BRmetric}
\end{equation}

The last term is the extra contribution to the perturbation arising from a
gauge fixing different from the DeWitt one. For example, for the 
$\lambda$ family we have $a(\lambda)=-\frac{5 \lambda \kappa^2}{48 \pi^2}$ and 
$b(\lambda)=\frac{5 \lambda \kappa^2}{96 \pi^2}$. {\it It is then clear 
that the 
metric 
that solves the backreaction equations for the one loop quantized gravity 
depends on which particular function one chooses to fix the gauge.} 

\section{Quantum corrected geodesic equation: Newtonian limit}

The dependence on the gauge fixing of the 
gravitons is an
obstacle to think of a
solution to the SEE as the metric of spacetime. 
This obstacle is not
`technical' (as implicitly assumed in previous works) but physical: since any
classical device couples to
gravitons, the solution to the SEE will not, in general,
have a clear physical interpretation. 

To analyze this problem, we will
consider  the simplest classical device: a  test particle 
of mass $m$ in
the presence of the quantized gravitational field 
${\bar g}_{\mu\nu}$. A physical observable should be the motion of 
this particle. 
We consider that the mass of this particle is much smaller than $M$, which 
allows us to neglect all contributions of the test particle to the solution
of the one loop corrected equation (\ref{eq:BRmetric}).
Now comes the key ingredient: in order to determine how this test 
particle moves, one also has to take into account the fact that it couples to 
the quantum metric ${\bar g}_{\mu\nu}$ through a term 
$-m \int \sqrt{- {\bar g}_{\mu\nu}(x) dx^{\mu} dx^{\nu} }$, where $x^{\mu}$ 
denotes the path of the test particle. Therefore there will be an extra 
contribution to the one loop EA due to this coupling to gravitons, which
in turn will introduce a correction to the geodesic equation. This contribution
is, up to linear order in $m$,

\begin{eqnarray}
\Delta S_m &=&  \int d^4x \sqrt{-g}
\left[ 
-\frac{1}{64 \pi^2} m_{\mu\nu\rho\sigma} \ln(-\Box) M^{\rho\sigma\mu\nu} -
\frac{1}{32 \pi^2}  m_{\mu\nu\rho\sigma} \ln(-\Box) 
   \left( P^{\rho\sigma\mu\nu} -
       \frac{1}{6} R \delta^{\rho(\mu} \delta^{\sigma\nu)} \right) +
\right. \nonumber\\
&& \left. ~~~~~~~~~~~~~~~~~~~~~~~~~~~~~~~~~~~~~~~~~~~~~~~~~~~~~~~~~~~~~
\frac{a}{2} R_{\mu\nu} \ln(-\Box) T^{\mu\nu}_m + 
   \frac{b}{2} R g_{\mu\nu} \ln(-\Box) T^{\mu\nu}_m \right],
\end{eqnarray}
where the tensor $m_{\mu\nu\rho\sigma}$ is the one given in Eq.(\ref{eq:meff})
with $M$ replaced by $m$, and $T^{\mu\nu}_m$ is the energy-momentum tensor
for the test particle, given in Eq.(\ref{eq:tmunu}), with the same replacement.
The first two terms correspond to the $\lambda=0$ gauge fixing, and the last
two are extra terms appearing for any other gauge. In the weak, 
nonrelativistic Newtonian limit, the quantum corrected geodesic equation reads

\begin{equation}
\frac{d^2{\vec x}}{dt^2} - \frac{1}{2} {\vec \nabla}h_{00} =
\frac{1}{m} \frac{\delta \Delta S_m}{\delta {\vec x}} .
\end{equation}

Note that $h_{00}$, given in Eq. (\ref{eq:BRmetric}), depends on $a$ and
$b$. The term on the rhs can be computed following the same methods we used
to solve the backreaction problem. In that way 

\begin{equation}
\frac{\delta \Delta S_m}{\delta {\vec x}} = 
\left[ \frac{5 G}{12 \pi} - a + 2b \right]  
{\vec \nabla} \left( \frac{G m M}{r^3} \right) .
\end{equation} 

Plugging this expression into the corrected geodesic equation we see that
those gauge fixing dependent terms arising from the backreaction metric 
cancel exactly those coming from the coupling of the test particle to
gravitons. In this way we obtain a physical, gauge fixing independent 
Newtonian potential $V(r)$ which we read from 
$d^2 {\vec x}/dt^2 = - {\vec \nabla} V$, namely
   
\begin{equation}
V(r) = -\frac{G M}{r} \left[1 + \frac{43 G\hbar}{30 \pi r^2 c^3} - 
\frac{7 G \hbar}{12 \pi r^2 c^3} +\frac{5 G \hbar}{12 \pi r^2 c^3} \right] ,
\end{equation}
where we have restored units ($\hbar$ and $c$). Note that the long distance
quantum correction above is extremely small 
to be measured. {\it However the
specific number is less important than the conceptual fact that the potential
and motion of the test particle
are gauge fixing independent}. 

\section{Cosmological background geometries}

Up to here we have considered the quantum corrections to the metric and
to the test particle trajectory in the Newtonian approximation. We will
now briefly comment about the quantum corrections to the geodesics in
a cosmological background. 

For simplicity, instead of working with a 
general gauge fixing term (as we did before), we will fix completely
the gauge of the quantum fluctuations and describe gravitons by two
massless,
minimally coupled scalar fields. Moreover, we will
assume that we are able to obtain a cosmological  solution to the SEE,
including graviton fluctuations, and will focus only in the computation 
of quantum
corrections to the geodesic equation. 
The solution to the SEE will be described by the line
element $
ds^2=-dt^2+a^2(t)(dx^2+dy^2+dz^2)$ for some function $a(t)$. 

The corrections to the
geodesic equation can be computed following a procedure similar to the 
one described in the previous section. The coupling between the gravitons
and the test particle modifies the classical action for the particle.
Keeping only the corrections that are linear in the mass $m$ the
corrected geodesic equation reads

\begin{eqnarray}
 && {d^2t\over d\tau^2} + a \dot a \left({d\vec x\over d\tau}\right)^2= 4\pi G
{dx^j\over d\tau}{dx^k\over d\tau}{dx^l\over d\tau}{dx^m\over d\tau}
{\partial\over\partial t}G_{jklm}\nonumber\\
&& {d\over d\tau}[a^2 {d\vec x\over d\tau}]=16 \pi G{d\over d\tau}
[G^n_{jkl}{dx^j\over d\tau}{dx^k\over d\tau}{dx^l\over d\tau}]
\end{eqnarray}
Here a dot denotes derivative with respect to $t$. $G_{ijkl}$   is  the (renormalized)
coincidence limit of the graviton two point funtion $<h_{ij}(x)h_{kl}(x')>$. 
Note that $G_{ijkl}$ only depends on $t$. 

Taking into account the symmetries of the problem, the above equations can be easily 
solved. 
Let us assume that, when the graviton contribution is neglected, the particle moves 
in the $x$ direction
($x=x^1$). Hence, in this case,

\begin{eqnarray}
  {d x \over d\tau} &=& {\alpha\over a^2}\nonumber\\
{d t \over d\tau} &=& \sqrt{1+{\alpha^2\over a^2}}
\end{eqnarray}
 where $\alpha$ is a constant related to the velocity of the test particle. 
Note that $\alpha\rightarrow \infty$ in the null limit. The above equation
describes, of course, the classical trajectory of the particle.

After a strightforward calculation we obtain, to first order in the quantum 
corrections,

\begin{eqnarray}
  {d x \over d\tau} &=& {\alpha\over a^2}
\left[  1+{32\over 3}\pi G <\varphi ^2> {\alpha^2\over a^2} \right]
\nonumber\\
{d t \over d\tau} &=& \sqrt{1+{\alpha^2\over a^2}}
\left[ 1+{32\over 3}\pi G <\varphi ^2> 
{{\alpha^4\over a^4}\over 1 + {\alpha^2\over a^2}  }\right]
\end{eqnarray}
where $<\varphi^2>$ is the renormalized coincidence limit 
of the two point function
for a massless, minimally coupled scalar field in the background described by
$a(t)$.

From    
the above equation we can find the quantum correction to the physical 
velocity of the test
particle, induced by its coupling to the gravitons,

\begin{equation}
{dx\over dt}={\alpha a^{-2}\over\sqrt{1 + \alpha^2 a^{-2}}}
\left[ 1+
   {32\over 3}\pi G <\varphi ^2> {{\alpha^2\over a^2}\over 1 + {\alpha^2\over a^2}  }\right]
\end{equation}

In the null limit ${dx\over dt}\simeq {1\over a} [1+{32\over 3}\pi G 
<\varphi ^2>] $
describes the graviton correction to the cosmological redshift.
We expect this 
correction to be  very small for $t\gg t_{Planck}$.

Had we considered a different gauge fixing, we would have obtained a different
expression for the quantum corrections to the geodesic equation in a
cosmological context. However, 
the solution to the SEE described by
$a(t)$ would also depend on the gauge fixing. As in the case of the
Newtonian potential, both dependences 
should cancel when computing the trajectory of the test 
particle.

\section{Conclusions}

We hope to have convinced you that if one is interested in
solving the backreaction problem including the graviton contribution,
{\it it is not enough to solve the semiclassical Einstein equations.
The solutions are gauge fixing dependent and not physical.} Rather one has to
look for physical observables. As an illustration of this point we have
shown, in the Newtonian approximation, that a classical test particle
does not follow a geodesic in the background metric. Moreover, 
the trajectory is a physical observable independent of the
gauge fixing procedure.
We have also computed the quantum corrected geodesic equation for a
test particle in a Robertson Walker background.

If the calculations presented here are performed using the
Vilkovisky-DeWitt effective action, we expect the dependence on the 
supermetric to 
cancel and the resulting quantum corrected geodesic 
equations to coincide with the ones
obtained by means of the conventional effective action.

Similar ideas to the one proposed here can be applied to more general
situations and even to the analysis of the mean value equations of any
gauge theory, for example when computing gluon backreaction effects
on classical solutions to Yang Mills theories.

The results presented in this talk are contained in Refs. \cite{nos1,nos2}.

\section{ Acknowledgements}

We acknowledge the support  
from Universidad de Buenos Aires, Fundaci\'on Antorchas and CONICET (Argentina).


\begin{references}
\bibitem{SIGNA}
Our metric has signature $(-+++)$ and the curvature tensor is defined as
${\bar R}^{\mu}_{\, \cdot \, \nu\alpha\beta} = 
\partial_{\alpha} \Gamma^{\mu}_{\nu\beta} - \ldots$, 
${\bar R}_{\alpha\beta} = {\bar R}^{\mu}_{\, \cdot \, \alpha\mu\beta}$ and
${\bar R}= {\bar g}^{\alpha\beta} {\bar R}_{\alpha\beta}$. We use units
$\hbar=c=1$.
\end{references}
\end{document}